\documentclass [prb,twocolumn,superscriptaddress] {revtex4}
\usepackage{graphicx}

\begin{document}

\title{Impact of dimerization and stretching on the transport
properties of molybdenum atomic wires}
\author{A. Garc\'ia-Fuente}
\affiliation{Departamento de F\a'{\i}sica Te\a'orica, At\a'omica y
\'Optica. Universidad de Valladolid, E-47011 Valladolid, Spain}
\author{A. Vega}
\affiliation{Departamento de F\a'{\i}sica Te\a'orica, At\a'omica y
\'Optica. Universidad de Valladolid, E-47011 Valladolid, Spain}
\author{V. M. Garc\'ia-Su\'arez}
\affiliation{Departamento de F\a'{\i}sica, Universidad de Oviedo \& CINN,
Spain}
\affiliation{Department of Physics, Lancaster University,
Lancaster, United Kingdom}
\author{J. Ferrer}
\affiliation{Departamento de F\a'{\i}sica, Universidad de Oviedo \& CINN,
Spain}

\date{\today}
\begin{abstract}
We study the electrical and transport properties of
monoatomic Mo wires with different structural characteristics. We
consider first periodic wires with inter-atomic distances ranging
between the dimerized wire to that formed by equidistant atoms. We
find that the dimerized case has a gap in the electronic structure
which makes it insulating, as opposed to the equidistant or
near-equidistant cases which are metallic. We also simulate two
conducting one-dimensional Mo electrodes separated by a scattering
region which contains a number of dimers between 1 and 6. The
$I-V$ characteristics strongly depend on the number of dimers and
vary from ohmic to tunneling, with the presence of different gaps.
We also find that stretched chains are ferromagnetic.
\end{abstract}



\maketitle

\section{Introduction}

The field of Nanoelectronics has emerged as one of the most
important branches of Nanoscience and Nanotechnology, due to both
the fundamental quantum phenomena occurring at this scale and to
its technological relevance in the context of miniaturization and
performance improvement of electronic devices. Remarkable progress
in the experimental techniques, achieved during the last years,
for growing, manipulating and measuring at the nanoscale has
allowed to demonstrate several prototype devices such as
conducting wires, point contacts and switches.\cite{nitzan}

Metallic atomic-scale nanowires, in particular, have attracted
considerable efforts devoted to their production and
isolation.\cite{benz,zach,nilius} Also infinite monoatomic
wires have been theoretically investigated recently in order to
better understand the magnetic and transport properties of
one-dimensional atomic contacts.\cite{autes,miura,tosatti}
>From such idealized models we can gain
valuable information which can be compared later with the
realistic systems to get insight in the influence of defects,
impurities, host material or the structure of the contacts.

Understanding the transport properties like the current flow
through these nanoscale devices, has been a challenge from the
theoretical side due to the non-equilibrium quantum kinetic
description required. The modern theory of quantum transport is
based on the scattering theory or on the Keldysh-Kadanoff-Baym
non-equilibrium Green's function (NEGF) formalism.\cite{datta} The
SMEAGOL code\cite{smeagol,towards} is a flexible and efficient
implementation of the NEGF formalism. SMEAGOL obtains the
Hamiltonian from the density functional theory
(DFT)\cite{kohn-sham} code SIESTA,\cite{siesta} which uses
pseudopotentials and a localized basis set of pseudo-atomic
orbitals, and calculates self-consistently the density matrix, the
transmission and the current for each bias voltage.

Very recently, individual molybdenum chains have been produced and
controlled by encapsulating them inside carbon
nanotubes.\cite{muramatsu} Mo is indeed an interesting metal
element as regards its ability to form one-dimensional (1D)
structures, even in free environment. Recent ab-initio
calculations of free-standing Mo atomic clusters have shown that
linear atomic chains containing up to four atoms are considerably more
stable than two- and three- dimensional structures\cite{faustino}.
This result is
relevant in the context of
Mechanically Controllable Break Junction experiments (MCBJE),
where atomic chains are fabricated at ultra-low temperatures by
bending and eventually breaking up an atomic strip. Interestingly,
only the elements gold, platinum and iridium have been found to
make them,\cite{smit} although other elements like Sn could be
possible candidates.\cite{lucas}. Due to its tendency to make
linear clusters\cite{faustino} Molybdenum could also be used to
make atomic chains in MCBJE. Therefore, a transport calculation of
such chains is indeed timely.

Besides, as a consequence of Mo having an exact half band filling,
the geometry is characterized by the formation of tightly bound
dimers, which lead to a insulating ground state. Further, these
chains show a Peierls-type insulator-metal which is activated
when the two inter-atomic distances are varied so that the dimerization
is lost. This article analyzes the impact of this metal insulator
transition in the transport properties of the chains.

More specifically, we study the electrical and transport properties of
infinite monoatomic Mo wires with different structural
characteristics. We consider first periodic wires with
inter-atomic distances ranging between the dimerized wire to that
formed by equidistant atoms. We have also simulated two conducting
one-dimensional Mo electrodes separated by a scattering region
which contains a number of dimers between 1 and 6. Since the
conductance is closely related to the electronic structure, which
depends dramatically on the distance between atoms, interesting
effects are expected to appear as a function of the structural
changes induced by the dimerization.

In the next section we give the details of our DFT calculations
within the SMEAGOL code. In Section III we present the results
obtained for the infinite monoatomic Mo wires between the
dimerized and the equidistant configurations. Section IV is
devoted to the scattering region formed by finite dimerized chains
anchored to infinite wires with equidistant atoms. In section V we
discuss the results for a Mo wire with interatomic distances
similar to those experimentally observed when the wire is
encapsulated in the nanotubes. Magnetic effects are expected here
due to the considerable expansion of the interatomic distances.
Finally, Section VI summarizes the main conclusions.

\section{Details of the DFT approach}

\begin{figure}
\includegraphics[width=0.8\columnwidth]{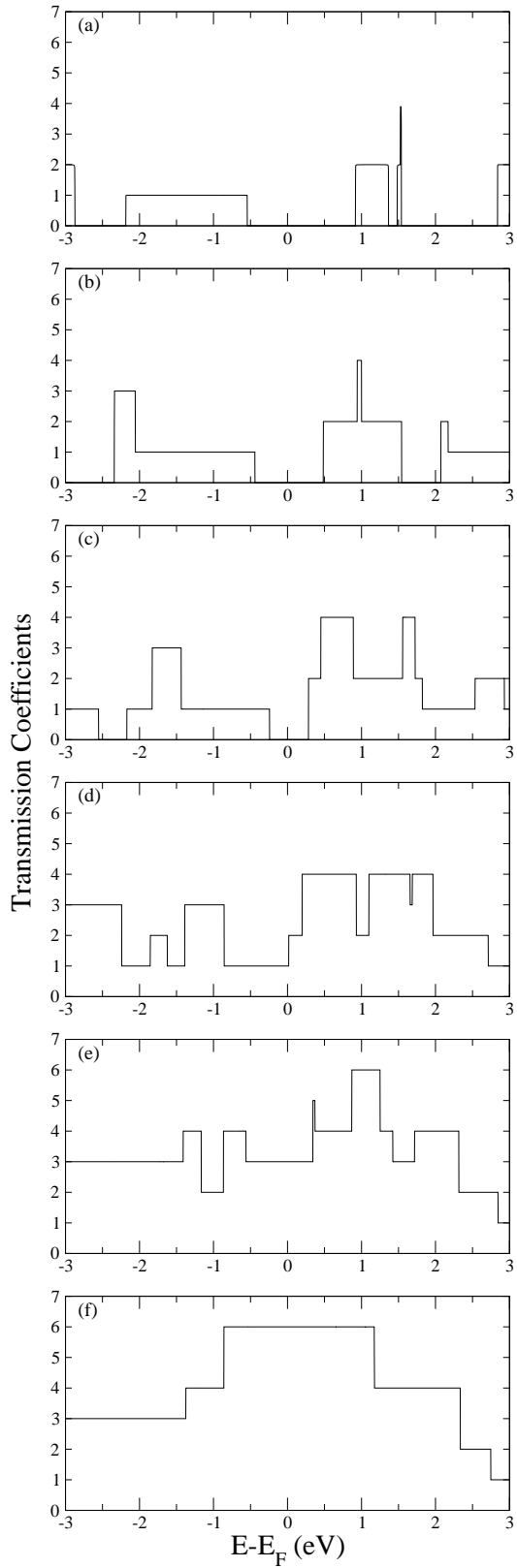}
\caption{Transmission as a function of the energy referred to the
Fermi level for a Mo atomic wire with interatomic distances of (a)
1.58 \AA\ and 3.01 \AA\ (b) 1.74 \AA\ and 2.85 \AA\ (c) 1.88 \AA\
and 2.71 \AA\ (d) 2.02 \AA\ and 2.57 \AA\ (e) 2.16 \AA\ and 2.43
\AA\ (f) equidistant 2.30 \AA} \label{Figure1}
\end{figure}

We calculated the exchange and correlation potential with the
generalized gradient approximation (GGA) as parametrized by
Perdew, Burke and Ernzerhof.\cite{PBE} We replaced the atomic Mo
core by a nonlocal norm-conserving Troullier-Martins\cite{TM_1991}
pseudopotential which was factorized in the Kleinman-Bilander
form\cite{KB_1982} and generated using the atomic configuration
$4d^{5}$ $5s^1$ $5p^0$, with cutoff radii of $1.67$, $2.30$ and
2.46 a.u., respectively. We also included nonlinear core
corrections, generated with a radius of 1.2 a.u, to account for
the significant overlap of the core charge with the valence d
orbitals and avoid spikes which often appear close to the nucleus
when the GGA approximation is used. We tested that this
pseudopotential reproduced accurately the eigenvalues of different
excited states of the isolated Mo atom. We employed a linear
combination of pseudoatomic orbitals to describe the valence
states of our chains.\cite{siesta} The basis set included
double-$\zeta$ polarized (DZP) orbitals, i.e. two radial functions
to describe the $5s$ shell and another two for the $4d$ shell,
plus a single radial function for the empty p shell. We used a
numerical grid defined by an energy cutoff of 250 Rydberg to
compute the exchange and correlation potential, and to perform the
real-space integrals that yield the Hamiltonian and overlap matrix
elements. We also smoothed the Fermi distribution function that
enters in the calculation of the density matrix with an electronic
temperature of 300 K and used a conjugate gradients
algorithm,\cite{NR} to relax the atomic positions until the
interatomic forces were smaller than 0.01 eV/\AA. Finally, we
performed careful tests for particular cases to ensure the quality
and stability of the basis set and the real space energy cutoff
employed. We found that the results were hardly modified when the
DZP basis was replaced by a triple-$\zeta$ doubly polarized basis.
Similar results were also obtained by considering an electronic
temperature of 100 K.

\section{Infinite periodic monoatomic $\textbf{Mo}$ wires}

\begin{figure}
\includegraphics[width=1.0\columnwidth]{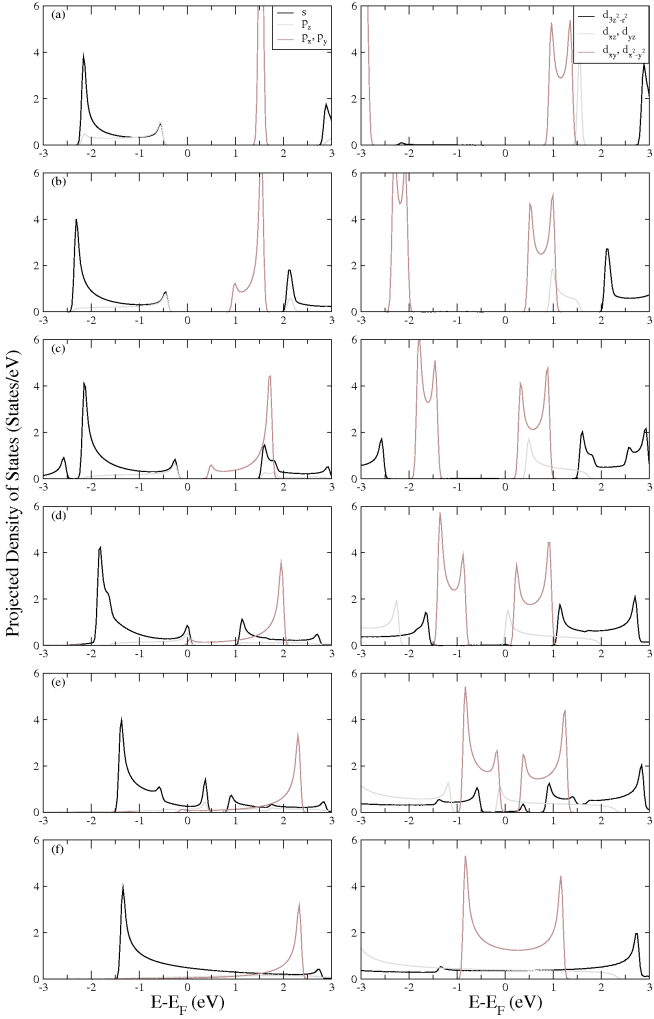}
\caption{Density of States projected on the hybrid cartesian s-,
p- (first column) and d- (second column) orbitals, as a function
of the energy referred to the Fermi level for the same wire
configurations as those of Fig. (\ref{Figure1}).} \label{Figure2}
\end{figure}

The ground state of the periodic monoatomic Mo wire is
formed by tightly bonded dimers of 1.58 \AA\ inter-atomic distance
and 3.01 \AA\ inter-dimer distance. The resulting wire is
non-magnetic, since the Mo atom has exactly a half filled d-shell,
which is a consequence of the strong covalent bond formed between
the two Mo atoms of the dimer to achieve a closed-shell electronic
configuration. Notice that dimer formation arises in finite length
chains containing as few as 4, 6 and 8 atoms, where one can not
strictly speak of electronic bands. This phenomenon is therefore
a Peierls dimerization transition driven by the electron-phonon
coupling.\cite{faustino}
Indeed, the magnetic nature of these chains is paramagnetic,
as opposed to the antiferromagnetic character which would result
by using nesting arguments. Further, chains with an odd number of
Mo atoms (3, 5, 7) also dimerize, the dimers being in a singlet
state, and the unpaired atom providing $5\mu_B$.

These chains are electronic insulators. To further qualify their
insulating behavior, we have used SMEAGOL to compute the
transmission coefficients. SMEAGOL computes the current using the
Landauer formula\cite{datta}

\begin{equation}
I(V)\,=\,\frac{2\,e}{h}\, \int \,\mathrm{d}E\, T(E,V)\,\left(
f_L(E,V)-f_R(E,V)\right)
\end{equation}

\begin{figure}
\includegraphics[width=1.0\columnwidth]{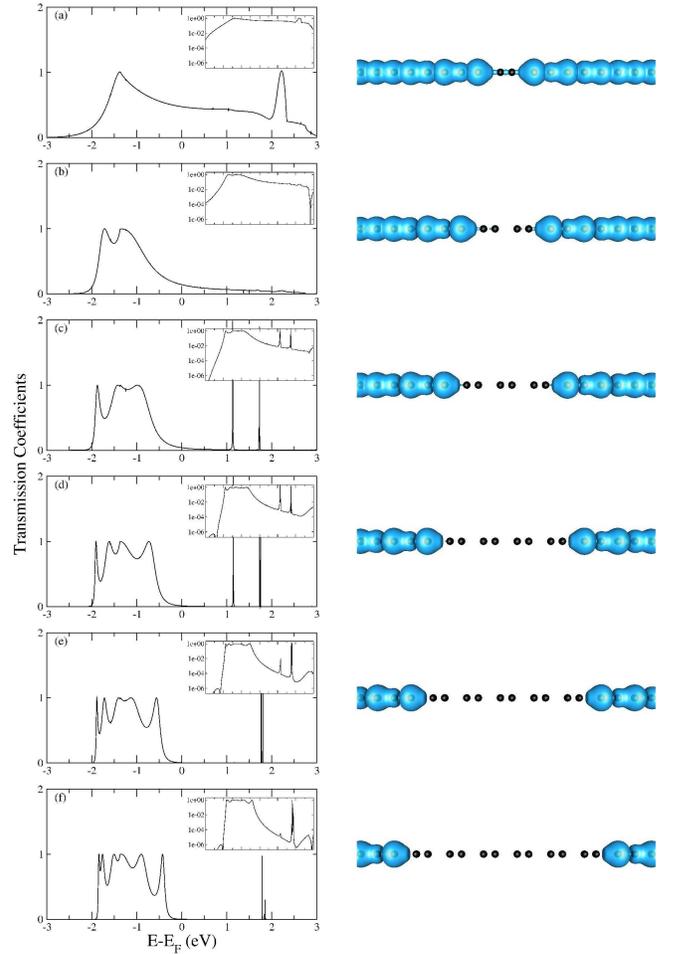}
\caption{(color online) Transmission (first column) and real space
plot of the charge density around the Fermi level (second column)
for a scattering region with a number of dimers ranging from 1 to
6.} \label{Figure3}
\end{figure}

\noindent where $T(E,V)$ are the energy- and voltage-dependent
transmission coefficients of the junction, and
$f_\mathrm{L/R}(E,V)=f(E-\mu_\mathrm{L/R})$ are the Fermi
functions of the left/right electrodes, whose chemical potentials
$\mu_\mathrm{L/R}=\mu \pm eV/2$ are the equilibrium chemical
potential shifted by the voltage bias $V$. The low-energy and
low-voltage transmission coefficients can be expanded as

\begin{equation}
 T(E,V) \simeq T_0,+T_E\,E+T_V\,V + O(E^2,V^2)
\end{equation}

\noindent where $T_0=T(0,0)$, and $T_{E,V}=\frac{\partial
T}{\partial (E,V)}(E=0,V=0)$. As a consequence, the low-voltage
differential conductance is equal to

\begin{equation}
G(V)=\frac{\mathrm{d}I}{\mathrm{d}V}\simeq G_0
\,\left(T(0,0)+2T_V\,V\right)\simeq G_0 T(0,0)
\end{equation}

\noindent where $G_0=2\,e^2/h$ is the conductance quantum unit.
Notice that the first-order energy term in the Taylor expansion
drops because of the symmetry properties of the integral.
Therefore, the zero-energy, zero-voltage transmission coefficients
provide an estimate of the low-voltage differential conductance.
Furthermore, for symmetric junctions such as those that will be
studied in this article $T(E,V)=T(E,-V)$, so that $G(V)=G(-V)$.
The differential conductance must be flat at very low voltages,
and only second-order voltage terms  contribute.

\begin{figure}
\includegraphics[width=0.7\columnwidth]{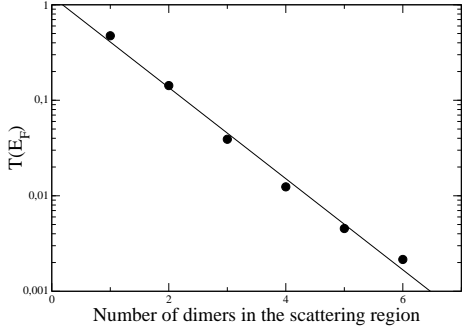}
\caption{Logarithm of the transmission at the Fermi level as a
function of the number of dimers in the scattering region.}
\label{Figure4}
\end{figure}

We plot in Fig. (\ref{Figure1}) (a) the zero-bias transmission as
a function of energy $T(E,0)$ for the dimerized infinite Mo atomic
wire. This chain does not have conduction channels in a window of
about 1.5 eV around the Fermi energy, which is confirmed by
plotting the Density of States (DOS) $N(E)$.
Notice that in a ballistic one-dimensional conduction model the
zero-voltage transmission coeficient at a given energy can be
written as

\begin{equation}
 T(E)\propto \sum_n v_n(E) N_n(E)
\end{equation}

\begin{figure}
\includegraphics[width=0.8\columnwidth]{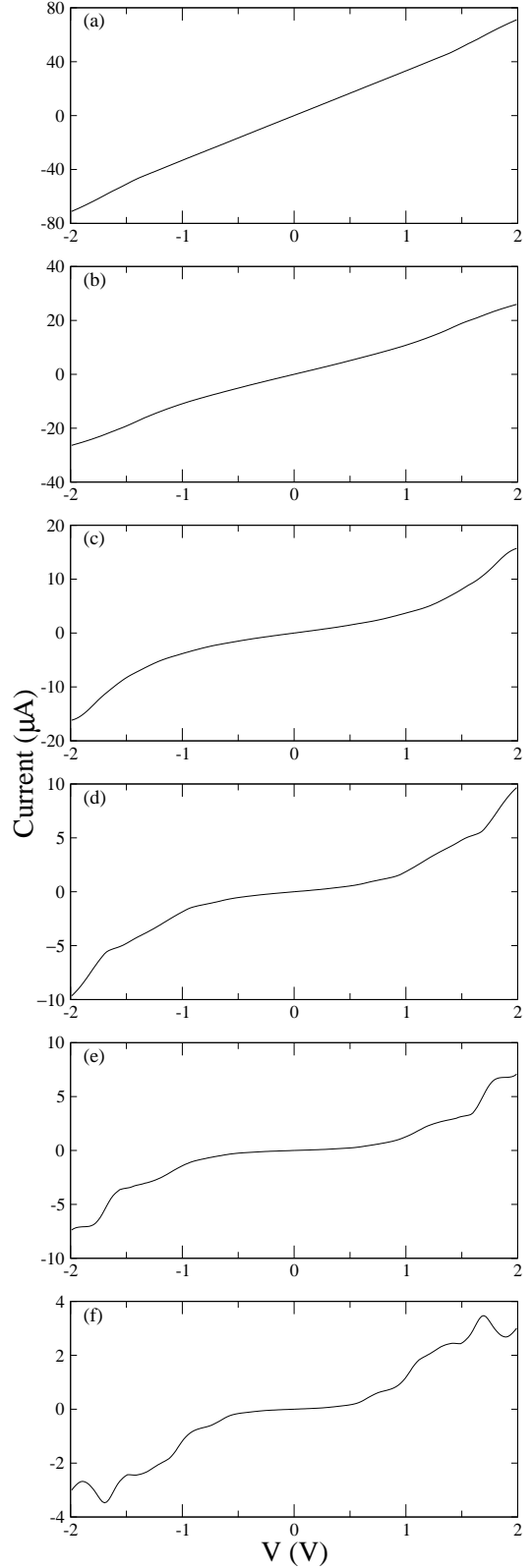}
\caption{$I-V$ characteristics for a chain with a scattering
region with a number of dimers ranging from 1 to 6.}
\label{Figure5}
\end{figure}

\noindent where the sum is over all the electron bands, and
$v_n(E)$ and $N_n(E)$ are the group velocity and the density of
states at energy $E$ of each band. Since in one dimension,
$N_n=1/(\hbar v_n)$, the transmission coefficient $T(E)$ just
counts the number of bands at energy $E$, which are called
transmission channels. For energies below -0.5 eV the transmission
exhibits a plateau with a magnitude of 1, which corresponds to a
single low-scattering sp band, or channel, and for energies above
1 eV, $T=2$ due to contributions from the antibonding d orbitals.

\begin{figure}
\includegraphics[width=0.8\columnwidth]{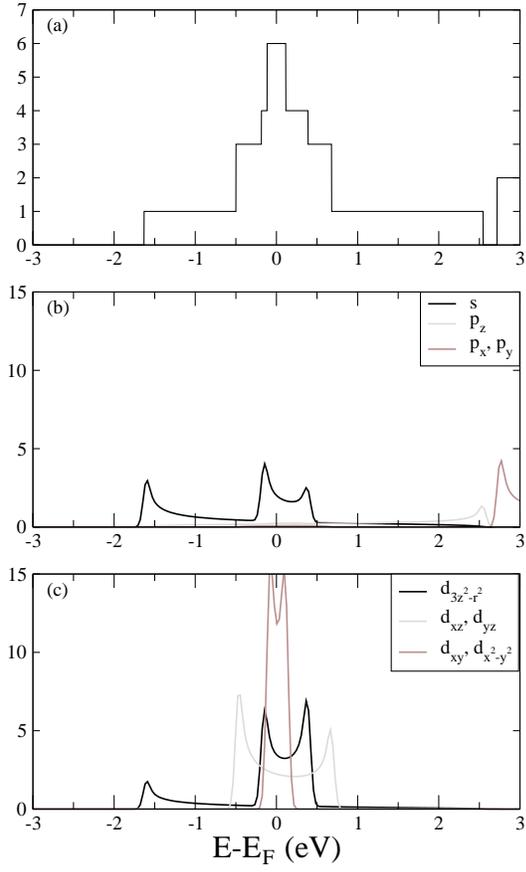}
\caption{Transmission (a) and projected density of states  on the
hybrid cartesian s-, p- (b) and d- (c) orbitals, for a Mo chain
with an interatomic distance of 3.5 \AA. The energies are referred
to the Fermi level.} \label{Figure7}
\end{figure}

In order to understand the interplay between dimerization and
transport, we have calculated the zero-voltage transmission
coefficients $T(E)$ for six chains with different inter-atomic
distances. The unit cell of all chains contains two atoms, and has
a lattice constant $a=4.6$ \AA. We have varied the inter-atomic
distance $(d)$ of the two atoms in the unit cell from $d_1 =1.58$
\AA\ (ground state dimer) to $d_2=2.30$ \AA\ (equidistant atoms).
To characterize all these chains, we introduce the dimer order
parameter $\Delta=2 d/a$, which varies from $\Delta= 0.69$ for
$d=d_1$ to $\Delta=1$ for $d=d_2$.

We note first that the ground state of all chains is non-magnetic.
Figs. (\ref{Figure1}) (a) to (f) illustrate the evolution of the
transmission coefficients with the order parameter. What we find
is that as $\Delta$ increases from 0.69, many occupied and
unoccupied channels move gradually towards the Fermi level,
increasing the conductive nature of the chains. There is an
insulator-metal transition at about $\Delta=0.88$ (Fig (\ref{Figure1})
(d)) and for larger order parameters the chains become metallic.
In other words, the insulating character is retained for dimerized chains, where the
dimer is strained [$\Delta\in (0.69,0.88)$]. Conversely, not only
perfect ($\Delta=1$), but also slightly dimerized chains
[$\Delta\in (0.88,1)$] are metallic. Notice also that the width of
the electronic bands and the plateaus of $T(E)$ increases with
$\Delta$, indicating that the electronic states become more
delocalized.  The actual conductance $G=T(0)\,G_0$ increases
step-wise from $G_0$ at $\Delta=0.88$ to $6G_0$ at $\Delta=1$.

\begin{figure}
\includegraphics[width=0.8\columnwidth]{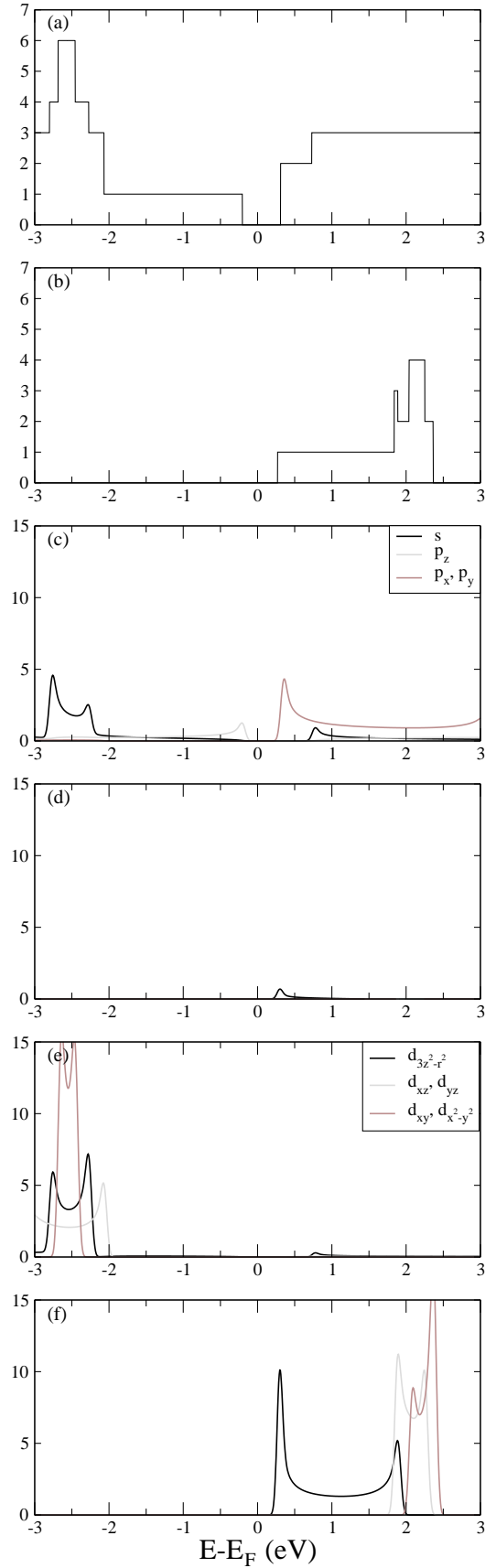}
\caption{Transmission and projected density of states on the
hybrid cartesian s-, p- and d- orbitals for the up (a) (c) (e) and
down (b) (d) (f) spin contributions, for a magnetic Mo chain with
an interatomic distance of 3.5 \AA. The energies are referred to
the Fermi level.} \label{Figure8}
\end{figure}

To understand how each state participates in the
conductance, we plot the evolution of the projected density of
states for each s-, p- and d-orbital in Fig. (\ref{Figure2}).
Notice the chains are oriented along the $z$-axis, which is also the
transport direction in our simulations. The d$_{3z^2-r^2}$
orbitals have therefore the largest hybridizations (that is to
say, Hamiltonian matrix elements) among all d-orbitals, followed
by the d$_{zx}$ and d$_{zy}$ orbitals (which are equivalent by
symmetry), leaving the d$_{x^2-y^2}$ and d$_{xy}$ as the less
hybridized orbitals. As a consequence, the d$_{3z^2-r^2}$ orbital
feels more strongly the dimerization of the chain, while those
states related to the d$_{x^2-y^2}$ and d$_{xy}$ orbitals are the
least affected. The most relevant effect of the dimerization is
the opening of a gap between the bonding and antibonding states,
which is largest for the  d$_{3z^2-r^2}$ orbital (6 eV for
$\Delta=0.69$) and smallest for the d$_{x^2-y^2}$ and d$_{xy}$
orbitals (3.5 eV for $\Delta=0.69$), while the gap is of 5 eV for
the states originated from the d$_{xz}$ and d$_{yz}$ orbitals.
There is also a hybridized sp$_{z}$ orbital below the Fermi
level, and a p$_{x}$ and p$_{y}$ contribution hybridized with the
d$_{xz}$ and d$_{yz}$  orbitals. When the order parameter $\Delta$
approaches 1, the first channel which reaches the Fermi level and
participates in the zero-bias conductance is the sp$_{z}$,
followed immediately by the d$_{xz}$ and d$_{yz}$. In contrast,
the gap belonging to the d$_{3z^2-r^2}$ orbital only closes when
$\Delta\sim 1$, and therefore only contributes to the zero-bias
conductance when the atoms in the chain are almost equidistant.
When the chain is completely equidistant the gaps in the densities
of states disappear and all the d-orbitals contribute to the
conductance. Thus, the 6 channels of conductance at the Fermi
level for the equidistant chain comes from the 5 d-orbitals and
from the sp$_{z}$-orbital, since the contribution of the p$_{x}$
and p$_{y}$ in the d-orbitals is negligible.

\section{One-dimensional $\textbf{Mo}$ electrodes separated by a
dimerized region}

In order to investigate the role of dimers in the scattering
region between two conductive electrodes we have considered the
formation of 1 to 6 dimers within an infinite equidistant wire.
Thus, at both sides of the scattering region we have ideal
conductive 1D electrodes. The question that arises is if the
current flow is disrupted and how by the dimers. The scattering
region, which we call the extended molecule, includes also part of
the electrodes (four atoms at each side) to ensure the continuity
of the potential with that of the infinite electrodes. To
determine the $I-V$ curve we need to evaluate the charge density
by using the lesser Green's function (instead of the retarded one,
since electrodes at different potential establish a nonequilibrium
population in the extended molecule) and the self-energies of the
electrodes at different potentials for all the energy range in the
bias window (not only at $E_\mathrm{F}$). The intensity of the
current at a given bias is obtained by using Eq. (1), where

\begin{equation}
T(E,V)=\mathrm{tr}[\Gamma_\mathrm{L}
G^{\mathrm{R}\,\dagger}_\mathrm{M}\Gamma_\mathrm{R}
G^\mathrm{R}_\mathrm{M}](E,V).
\end{equation}

\noindent $G^\mathrm{R}_\mathrm{M}=[\varepsilon^+
S_\mathrm{M}-H_\mathrm{M}-\Sigma_\mathrm{L}^\mathrm{R}-
\Sigma_\mathrm{R}^\mathrm{R}]^{-1}$ is the retarded Green's
function of the extended molecule, with
$\varepsilon^+=\mathrm{lim}_{\delta \to 0^+}E+i\delta$ and
$\Sigma_\mathrm{L(R)}^\mathrm{R}$ the retarded self-energy of the
left (right) electrode, and $\Gamma_\mathrm{L(R)}=
i[\Sigma_\mathrm{R(L)}^\mathrm{R}-\Sigma_\mathrm{L(R)}
^{\mathrm{R}\,\dagger}]$ is the Gamma matrix, which gives the
strength of the coupling between the scattering region and the
left (right) electrode.

We plot in Fig. (\ref{Figure3}) the zero-voltage transmission for
different scattering regions with an increasing number of dimers,
from 1 to 6. Each panel has attached to the right a real-space
plot of the density of states integrated in a window of 0.5 eV
about the Fermi energy. We find that, even if we introduce only
one dimer, the transmission at the Fermi level falls from 6 to
only 1 G$_0$, due to the lack of transmission channels in the
scattering region. For only one dimer we find that some electronic
density spreads around the dimer, but becomes rapidly localized
when we increase the number of dimers. This indicates again that
the transport must be in the tunneling regime, where the low bias
conductance decreases exponentially as a function of the distance
between electrodes, since electrons at the Fermi energy can only
hop from dimer to dimer via tunneling events. The transmission
$T(E)$ begins to resemble that obtained in the case of the
infinite dimerized wire (Fig. (\ref{Figure1}), top panel) when the
number of dimers increases.

To confirm the tunneling behavior of the conductance, we plot in
Fig. (\ref{Figure4}) the logarithm of the zero-voltage transmission
at the Fermi level in a logarithmic scale as a function of the number
of dimers in the scattering region. The points are indeed fitted by a linear
function, so the transmission coefficients fit to an exponential.
To find out the effective height $V_\mathrm{eff}$ of the tunneling barrier, we set

\begin{equation}
T(0,0)= A\, e^{-\sqrt{\frac{2\,m\,V_\mathrm{eff}}{\hbar^2}} \,d \,
N}= 1.224\, e^{-1.10 \,N}
\end{equation}

\noindent where $d=1.58$ \AA\ is the dimer length, $m$ is the
electron mass and $N$ is the number  of dimers in the chain.
Notice 
 that neither $m$ nor $V_{eff}$ are properly
defined separately. Rather, only their product $m\times V_{eff}$ can be extracted from the
fitting. We pick a common practice when fitting
conductance curves in vaccum tunneling, and select $m$ to be the
physical mass of the electron, thus dumping all the non-trivial
renormalizations into the effective height of the barrier. We
find that $V_\mathrm{eff}\sim 1.8$ eV, which agrees reasonably
well with the gap in the DOS of an infinite dimerized chain.

In Fig. (\ref{Figure5}) we plot the $I-V$ characteristics in a
voltage range between -2 and +2 volts, for the different
scattering regions mentioned above. We note that the $I-V$ curves
look like featureless Ohmic curves when the scattering region
comprises just one or two dimers. For a larger number of dimers
the $I-V$ curves start to develop a gap of about 2 volts, and a
number of additional high-voltage shoulders. These evolve into
negative differential resistance (NDR) characteristics for large
numbers of dimers. The NDR is produced by a small
peak in the transmission coefficients $T(E,V)$, which is located
around -0.75 eV. This peak appears for high
absolute values of $V$, but its height has a sudden drop when
the $V$ is increased beyond 1.7 volts. The drop leads to a
concomitant decreases in the electrical current. We find that
if $V$ is increased even further, a second peak enters inside
the integration window when $V\simeq 1.9$ volts, so that the
electrical current increases again. The emergence and evanescence
of transmission peaks can be understood in terms of the disorder
introduced by the bias-dependent on-site energies,\cite{Muj96}.
We have found a similar NDR effect in thiol-capped polyynes junctions.
\cite{Garciasu}


\section{Stretched monoatomic $\textbf{Mo}$ wires}

The interatomic distances $d$ recently measured between Mo atoms in
individual monoatomic chains encapsulated inside carbon
nanotubes\cite{muramatsu} ranged from 3.2 \AA\ to 3.8 \AA. Those
wires are then far from being dimerized; rather, they are stretched,
since those distances are much longer than 2.3 \AA,
which corresponds to equidistant free-standing chains.

We expect that isolated chains with such long $d$ become
ferromagnetic because of the Stoner criterion: they must have
narrower bands and, as a consequence, a high density of states
at the Fermi level in the non-magnetic state.
Indeed, we have found that the ground state configuration is
characterized by a full spin polarization, where the magnetic
moment is saturated to its maximum possible value of 6 $\mu_b$ per
atom, and by a large exchange splitting of xx eV. Hence,
all 5d as well as the 6s orbital are completely spin polarized.
This ferromagnetic ground state is more stable than the
paramagnetic state by 2.3 eV.

We expect that this spin polarization must have an
important impact in the transport behavior.
Indeed, H\"afner {\em et al.} \cite{Hafner}
found in a recent work that atomic contacts of Fe, Co and Ni carry spin polarized
currents, where the degree of the spin-polarization of the current
depended on the type of transport regime (contact or tunneling.
To estimate the extent of this impact, we have calculated the
transmission in both the non-magnetic and the magnetic states.

We have computed first the transmission coefficients of an isolated Mo wire in
the non-magnetic state with equidistant interatomic distances and a
unit cell similar to that found experimentally. The results are
shown in Fig. (\ref{Figure7}). We find that all six s- and d-channels
contribute to the transmission coefficient at the Fermi level, like in
the equidistant regime discussed in
section III. This might look surprising at first glance since
the distance between atoms in this stretched chain is even
larger than the inter-dimer distance of the dimerized chains with
insulating behavior discussed in section III. However, we note
that in the equidistant stretched chain the d-orbitals do not form
covalent bonds, and since no bonding and antibonding states are
formed (with their corresponding gap), the d-electrons are free to
participate in the transmission at the Fermi level. Away from the
Fermi level, however, the transmission coefficient decreases much faster than
in the previous case, since weak bonds due to electron
localization generate small bandwidths and, therefore, small
energy intervals in the transmission. The narrowing of the bandwidth
due to electron localization is clear if we compare the densities
of states of Fig. (\ref{Figure2}) with those corresponding to the
equidistant wire in Fig. (\ref{Figure7}) (lower panel).

We find that the ferromagnetic state is insulating. 	current is fully spin-polarized for the ferromagnetic
state, since all s and d channels with spin down (minority
electrons) between -3 and 3 eV contribute to the transmission
above to the Fermi level, as can be seen in Fig. (\ref{Figure8}).
We do not observe half-metallic behavior at the Fermi level,
however, due to the presence of the large spin splitting
which opens to completely occupy the d and s levels with unpaired
electrons. The encapsulation inside a nanotube is expected to
modify these electronic and transport properties, since there will
be a net charge transfer between the chain and the nanotube.

\section{Conclusions}

We have found that dimerization of molybdenum atomic wires has
dramatic effects on their electronic and transport properties.
While equidistant wires are metallic and have a very high
zero-bias conductance, dimerized wires show a large gap which
makes them insulating. As a consequence, the transport properties
of dimers between equidistant conducting wires show tunneling
behavior. The conductance decreases exponentially as the number of
dimers increases in the scattering region and the $I-V$
characteristics start developing a large gap around zero voltages.
Other features like plateaus and negative differential resistance
also appear for large numbers of dimers.

Equidistant stretched chains are highly conductive in the
non-magnetic case. However, the most stable configuration is
ferromagnetic, with a very high magnetic moment of 6 $\mu_B$ per
atom.

\acknowledgments This work was supported by the Spanish Ministry
of Education and Science in conjunction with the European Regional
Development Fund (Projects FIS2008-02490/FIS and FIs2006-12117),
and by Junta de Castilla y Le\'on (Project GR120).

{}

\end {document}